\documentclass[twocolumn,floatfix,aps,prb]{revtex4-2}
\usepackage{graphicx}
\usepackage{amsmath}
\usepackage{amssymb}
\usepackage{bm}
\usepackage{hyperref}
\usepackage{tikz}

\begin{document}

\title{Bloch oscillations in the magnetoconductance of twisted bilayer graphene}
\author{T. Vakhtel}
\affiliation{Instituut-Lorentz, Universiteit Leiden, P.O. Box 9506, 2300 RA Leiden, The Netherlands}
\author{D. O. Oriekhov}
\affiliation{Instituut-Lorentz, Universiteit Leiden, P.O. Box 9506, 2300 RA Leiden, The Netherlands}
\author{C. W. J. Beenakker}
\affiliation{Instituut-Lorentz, Universiteit Leiden, P.O. Box 9506, 2300 RA Leiden, The Netherlands}
\date{March 2022}
\begin{abstract}
We identify a mapping between two-dimensional (2D) electron transport in a minimally twisted graphene bilayer and a 1D quantum walk, where one spatial dimension plays the role of time. In this mapping a magnetic field $B$ perpendicular to the bilayer maps onto an electric field. Bloch oscillations due to the periodic motion in a 1D Bloch band can then be observed in purely {\sc dc} transport as magnetoconductance oscillations with periodicity set by the Bloch frequency.  
\end{abstract}
\maketitle

\textit{Introduction ---} 
It is one of the early counterintuitive predictions of solid state physics that an electric field in a crystal induces an \textit{oscillatory} electron motion \cite{Blo29,Zen34,history}: While the momentum $\hbar k$ increases linearly with time, according to $\hbar k(t)=e{\cal E}t$ in an electric field ${\cal E}$, the corresponding velocity $v(t)\propto \sin  k(t)$ in a Bloch band (unit lattice constant) has a periodic time dependence, with frequency $ \omega_{\rm B}=e{\cal E}/\hbar$. The amplitude ${\cal A}\approx\Delta/e{\cal E}$ of the Bloch oscillations is set by the energy band width $\Delta$.

Electronic Bloch oscillations have been studied in the time domain at THz frequencies in semiconductor superlattices \cite{Fel92,Leo92,Was93,Ros95,Kur96,Sav04} and in graphene bilayers \cite{Fah21}. With few exceptions \cite{Vil20}, and unlike the familiar Aharonov-Bohm oscillations \cite{Bee91}, Bloch oscillations do not typically play a role in quantum transport, which is probed in the energy domain at low frequencies. Here we show that Bloch oscillations may appear in the magnetoconductance of a two-dimensional (2D) system, a twisted graphene bilayer, by virtue of a mapping to a quantum walk in one space and one time dimension. 

The magnetic field $B$ perpendicular to the bilayer maps onto a parallel electric field ${\cal E}\leftrightarrow Bv/2$, with $v$ the Fermi velocity. As a consequence,  the conductance measured between two point contacts at a distance $L$ oscillates periodically in $B$. These Bloch magnetoconductance oscillations appear at much weaker fields, smaller by a factor $L$ over lattice constant, than the known Aharonov-Bohm oscillations in twisted bilayer graphene \cite{Ric18,Xu19,Beu20,Mah21}.

\begin{figure}[tb]
\centerline{\includegraphics[width=0.8\linewidth]{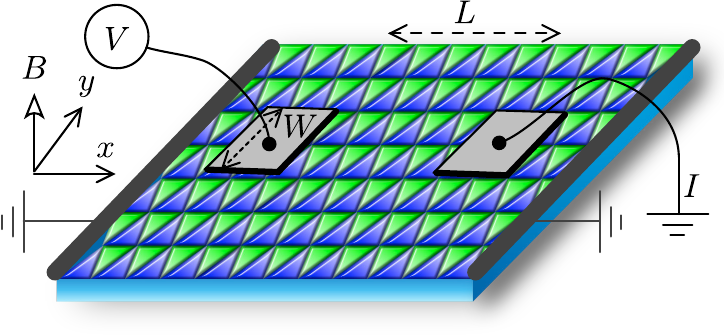}}
\caption{Moir\'{e} lattice in a twisted graphene bilayer. Triangular domains of AB and BA stacking are indicated by different colors. Domain walls conduct a current $I=VG$ between narrow source and drain contacts (width $W$, a distance $L$ apart). This is a four-terminal geometry, including two additional wide grounded contacts at the left and right.
}
\label{fig_bilayer}
\end{figure}

\textit{Network model ---}
We start from the established network model of minimally twisted bilayer graphene \cite{San13,Efi18,Hua18,Fle20,Tsi20,Beu21a,Ver21}: Two layers of graphene are misaligned by a rotation angle $\theta\approx 0.1^\circ$, forming a moir\'{e} pattern of triangular domains with different stacking (AB versus BA) of the carbon atoms on the A and B sublattices of the two layers. (See Fig.\ \ref{fig_bilayer}.) An interlayer bias voltage gaps out the interior of the AB and BA domains, leaving a conducting network formed by AB/BA domain walls that meet at angles of $60^\circ$ on a metallic node. The lattice constant $a=a_0[2\sin(\theta/2)]^{-1}$ of the moir\'{e} pattern is of the order of 100~nm, two orders of magnitude larger than the atomic lattice constant $a_0$ of graphene. 

The direction of motion along a domain wall is tied to the valley degree of freedom. (The spin degree of freedom is decoupled from the motion and plays no role in what follows.) In a single valley each domain wall supports two modes, both of the same chirality (propagating in the same direction with velocity $v$). Neglecting intervalley scattering (justified for $a\gg a_0$ with smooth disorder, and experimentally verified \cite{Ver21}) the scattering process at a node thus involves 6 incoming and 6 outgoing mode amplitudes, related by a scattering matrix $S$ of the form \cite{Beu20,Beu21a}:
\begin{subequations}
\label{Sdef}
\begin{align}
&S\cdot\{a_1,a_2,a_3,a'_1,a'_2,a'_3\}^\top=\{b_1,b_2,b_3,b'_1,b'_2,b'_3\}^\top,
\nonumber\\
&\begin{tikzpicture}[baseline=-0.5ex,scale=.5]
\node (O) at (0,0) {$$};
\node (A) at (2,0) {$a_1,a'_1$};
\node (B) at (-1,-1.73) {$a_3,a'_3$};
\node (C) at (-1,1.73) {$a_2,a'_2$};
\node (D) at (-2,0) {$b_1,b'_1$};
\node (E) at (1,1.73) {$b_3,b'_3$};
\node (F) at (1,-1.73) {$b_2,b'_2$};
\draw [double,->] (A) -- (O);
\draw [double,->] (B) -- (O);
\draw [double,->] (C) -- (O);
\draw [double,->] (O) -- (D);
\draw [double,->] (O) -- (E);
\draw [double,->] (O) -- (F);
\end{tikzpicture},\qquad 
S=\begin{pmatrix}
S_1&S_2\\
S_2^\dagger&-S_1^\dagger
\end{pmatrix},
\label{Sdefa}\\
&S_1=e^{i\alpha}\sqrt{P_{d1}}{\footnotesize{\begin{pmatrix}
0&1&1\\
1&0&1\\
1&1&0
\end{pmatrix}}}+e^{i\beta} \openone\sqrt{P_{f1}},\label{Sdefb}\\
&S_2=\sqrt{P_{d2}}{\footnotesize{\begin{pmatrix}
0&1&-1\\
-1&0&1\\
1&-1&0
\end{pmatrix}}}-\openone\sqrt{P_{f2}}.\label{Sdefc}
\end{align}
\end{subequations}

The $3\times 3$ submatrices $S_1$ and $S_2$ describe intramode and intermode scattering, respectively. Forward scattering happens with probability $P_f=P_{f1}+P_{f2}$, scattering with a $\pm 120^\circ$ deflection happens with probability $P_d=P_{d1}+P_{d2}$. Unitarity of $S$ requires that 
\begin{equation}
\begin{split}
&P_{f1}+P_{f2}+2P_{d1}+2P_{d2}=1,\\
&\cos(\beta-\alpha)=\tfrac{1}{2}(P_{d2}-P_{d1})(P_{f1}P_{d1})^{-1/2}\in[-1,1].
\end{split}
\end{equation}

To reduce the number of free parameters, we take equal intra-channel and inter-channel probabilities: $P_{f1}=P_{f2}=\tfrac{1}{2} P_f$ and $P_{d1}=P_{d2}=\tfrac{1}{4}(1-P_f)$. Then $\beta=\alpha+\pi/2$ and we are left with the two parameters $P_f\in[0,1]$ and $\alpha\in[0,\pi/2]$. The parameter $\alpha$ governs the appearance of closed loops of scattering sequences \cite{Beu20}. At $\alpha=0$, the network does not support closed loops (quasi-1D transport), while for $\alpha=\pi/2$ closed loops dominate (2D transport).

The propagation between two nodes along a domain wall introduces a phase factor $e^{iEa/\hbar v}$, at energy $E$. The scattering matrix \eqref{Sdef} of the nodes is assumed to be energy independent, so the scattering sequences consist of instantaneous nodal scattering events, spaced by the constant time $a/v$. A stroboscopic (Floquet) description of the scattering is then appropriate \cite{Cho20,Beu21b}. In the quasi-1D regime this corresponds to a quantum walk.

\begin{figure}[tb]
\centerline{\includegraphics[width=1\linewidth]{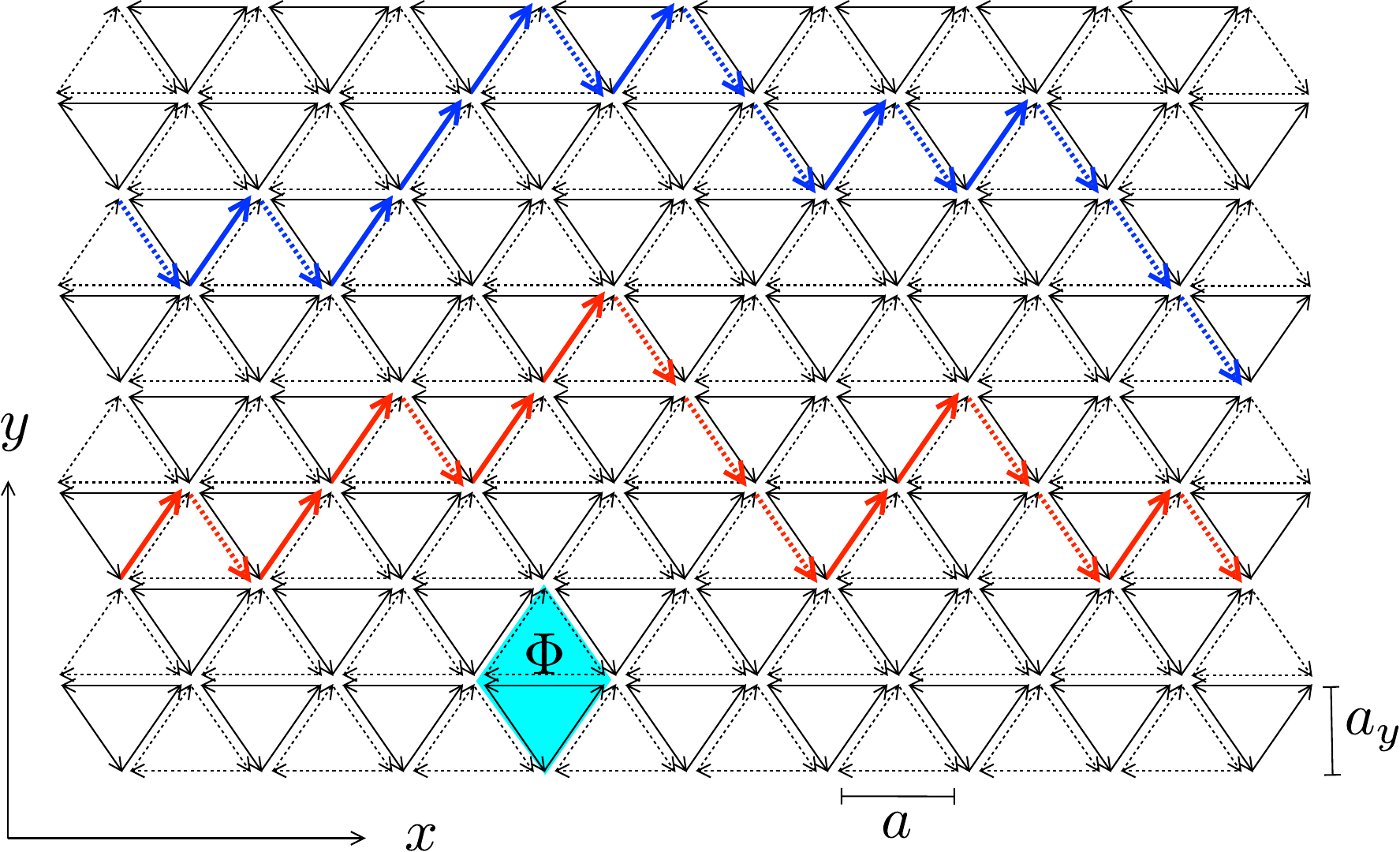}}
\caption{Network of domain walls with two realizations of the quantum walk in a single valley (blue and red arrows). Even and odd parity chiral modes are indicated by solid and dashed lines. Forward scattering at a node preserves the parity, while a deflection switches the parity. The quantum walk propagates along the $y$-axis with step size $a_y=\tfrac{1}{2}\sqrt{3}\,a$, the $x$-axis playing the role of time ($t\leftrightarrow 2x/v$). The unit cell of the lattice, of area $aa_y$ and enclosing a flux $\Phi$, is indicated in blue.
}
\label{fig_layout}
\end{figure}

\textit{Quantum walk ---}
Two scattering sequences in the quasi-1D regime ($\alpha=0$) are shown in Fig.\ \ref{fig_layout}. The solid and dashed lines distinguish even and odd parity modes in a given valley, both propagating in the same direction. (The counterpropagating modes are in the other valley.) We can interpret a scattering sequence as a quantum walk \cite{Kem03}, with time step $t_0=a/v$. There are six independent quantum walks, rotated relative to each other by $60^\circ$, three in one valley and three in the other valley. We focus on one of these.

At each step the $x$-coordinate is increased by $a/2$. The $y$-coordinate changes by $\pm \tfrac{1}{2}a\sqrt 3\equiv\pm a_y$, the even parity mode moves up and the odd parity mode moves down. The wave amplitudes $\psi=(\psi_+,\psi_-)$ of the even and odd parity states form a pseudospin degree of freedom, which is rotated at each node by the $2\times 2$ matrix \cite{Beu20,note1}
\begin{equation}
R=\begin{pmatrix}
e^{i\pi/4}\sqrt{P_f}&\sqrt{1-P_f}\\
\sqrt{1-P_f}&-e^{-i\pi/4}\sqrt{P_f}
\end{pmatrix}.\label{Rfdef}
\end{equation}

The corresponding time evolution of a state (at stroboscopic intervals $t=0,1,2,\ldots\times t_0$) is given by
\begin{align}
&\psi_{t+t_0}={\cal T}R\psi_{t},\label{RfTdef}\\
&{\cal T}\psi(y)=\bigl(\psi_+(y-a_y),\psi_-(y+a_y)\bigr)=e^{-ia_y\hat{k}_y\sigma_z}\psi(y).\nonumber
\end{align}
The operator ${\cal T}$ displaces the particle up or down depending on its pseudospin $\sigma_z$. Eq.\ \eqref{RfTdef} represents a 1D quantum walk along $y$ in the fictitious time $t=2x/v$, with momentum operator $\hat{k}_y=-i\partial/\partial y$.

\begin{figure}[tb]
\centerline{\includegraphics[width=0.6\linewidth]{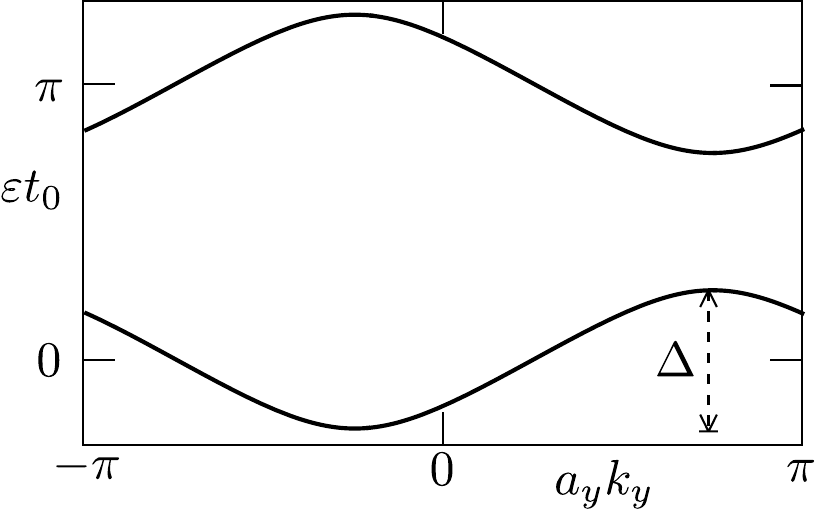}}
\caption{Two Bloch bands $\varepsilon_{\pm}(k_y)$ of the quantum walk, computed from Eq.\ \eqref{epspmn} for $P_f=1/2$. The band width $\Delta$ is indicated.
}
\label{fig_dispersion}
\end{figure}

The eigenvalues $e^{-i\varepsilon t_0}$ of the evolution operator ${\cal T}R$ in momentum representation are given by
\begin{equation}
\varepsilon_{\pm}t_0=\pm\arccos[\sqrt{P_f}\sin(a_yk_y-\pi/4)]+\pi/2,\label{epspmn}
\end{equation}
plotted in Fig.\ \ref{fig_dispersion}. The single-valley bandstructure of the quasi-1D regime \cite{Beu21a} is given by three copies rotated by $120^\circ$ of the dispersion relation 
\begin{equation}
E_{\pm}^{(n)}(k_x,k_y)=\hbar\varepsilon_{\pm}(k_y)+2\pi n\hbar/t_0+\hbar vk_x/2,\;\;n\in\mathbb{Z}.
\end{equation}

\textit{Bloch oscillations ---}
A perpendicular magnetic field $\bm{B}=\nabla\times\bm{A}$ (in the $z$-direction) introduces a phase shift $-e\int \bm{A}\cdot d\bm{l}$ at each time step (taking the electron charge as $+e$). For $\bm{A}=(-By,Ba/4,0)$ the time evolution \eqref{RfTdef} is modified into \cite{note2}
\begin{equation}
\psi_{t+t_0}=e^{i\phi\hat{y}/a_y}{\cal T}R\psi_t,\;\;\phi=\pi\Phi/\Phi_0.
\end{equation}
The operator $\hat{y}$ is defined by $\hat{y}\psi_t(y)=y\psi_t(y)$. The flux $\Phi=Baa_y$ is the flux through a unit cell (two domain wall triangles) and $\Phi_0=h/e$ is the flux quantum. The same phase shift $\phi$ would be produced by a fictitious electric field ${\cal E}\equiv Bv/2$. The corresponding Bloch frequency is 
\begin{equation}
\omega_{\rm B}=a_ye{\cal E}/\hbar=\phi/t_0.
\end{equation}
Since the width of the Bloch band \eqref{epspmn} is $\Delta=(2\hbar/t_0)\arcsin \sqrt{P_f}$, the amplitude of the Bloch oscillations is
\begin{equation}
{\cal A}\approx\Delta/e{\cal E}=(2a_y/\phi)\arcsin \sqrt{P_f}.
\end{equation}

\begin{figure}[tb]
\centerline{\includegraphics[width=1\linewidth]{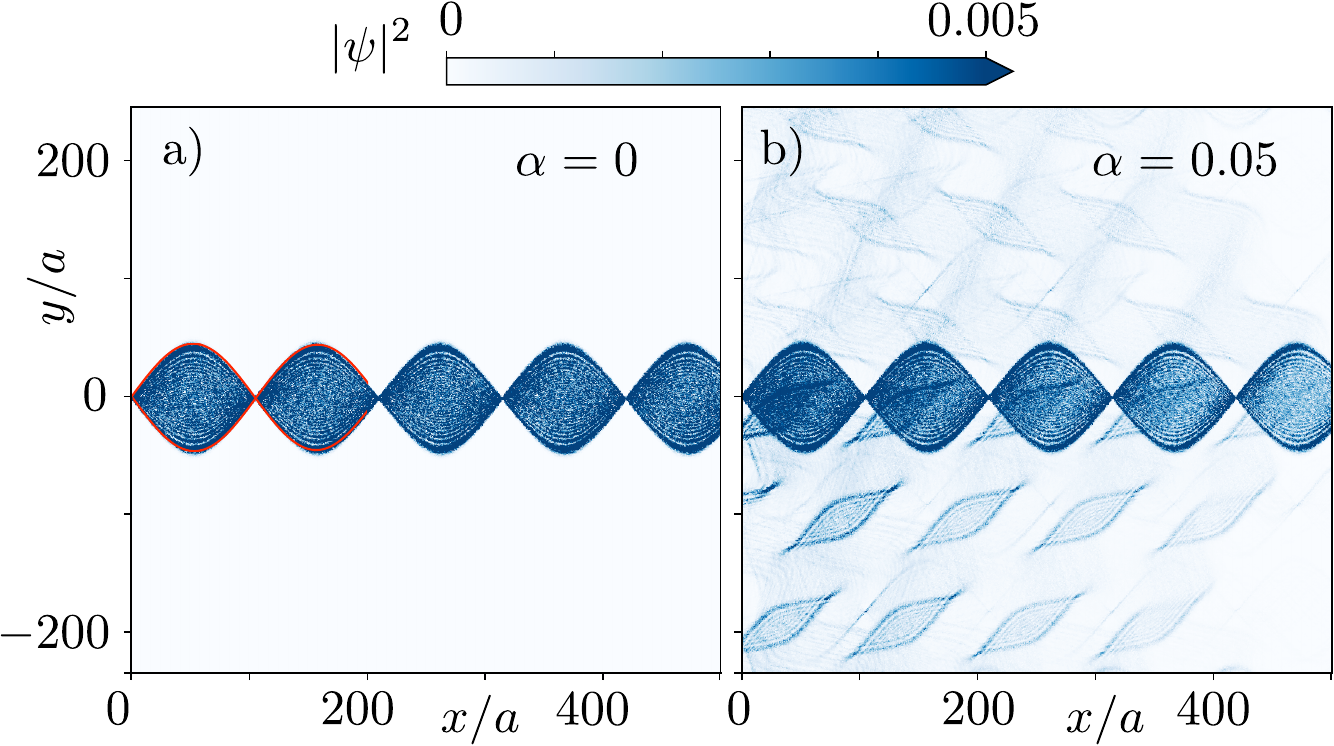}}
\caption{Numerical simulations of electron scattering in a twisted graphene bilayer, in a magnetic field ($\phi=\pi\Phi/\Phi_0=0.03$). Electrons at energy $E=0$ are injected in a single mode and in a single node at $(x,y)=(0,0)$. They then propagate through the network following the scattering matrix \eqref{Sdef}. We take $P_{f_1}=P_{f_2}=1/4$, $P_{d1}=P_{d2}=1/8$ and compare two values of $\alpha$. The blue color scale gives the intensity $|\psi|^2$ of the scattering state. The red curve in panel a) is the envelope of the breathing mode predicted by Eq.\ \eqref{yenvelope}.
}
\label{fig_breathing}
\end{figure}

The 1D quantum walk in an electric field has been analyzed theoretically \cite{Har04,Ced13,Arn20,Ced21} and realized experimentally in the context of optics \cite{Reg11,Err21} and atomic physics \cite{Gen13}. A spatially localized wave packet evolves in a characteristic ``breathing mode'' with envelope \cite{Har04} $\pm {\cal A}\sin(\omega_{\rm B}t/2)$. In our case, where time $t\mapsto 2x/v$ maps onto space, this implies the envelope
\begin{equation}
y_{\rm envelope}(x)=\pm (2a_y/\phi)\arcsin \sqrt{P_f}\times\sin(\phi x/a).\label{yenvelope}
\end{equation}

A numerical simulation of the network model shown in Fig.\ \ref{fig_breathing} agrees nicely with this breathing mode envelope. For nonzero $\alpha$ side branches appear at a $120^\circ$ orientation with the breathing mode, which we explain in terms of magnetic breakdown.

\textit{Magnetic breakdown ---} In semiclassical approximation the motion of an electron in a magnetic field $\bm{B}$ can be obtained from the equi-energy contours in zero field: Because $\hbar\dot{\bm{k}}= e\dot{\bm{r}} \times \bm{B}$, the real-space orbit of a wave packet at energy $E$ follows the contour $E(\bm{k})=E$ upon rotation by $90^\circ$ and rescaling by a factor $l_m^2=\hbar/eB$ (magnetic length squared).

\begin{figure}[tb]
\centerline{\includegraphics[width=0.6\linewidth]{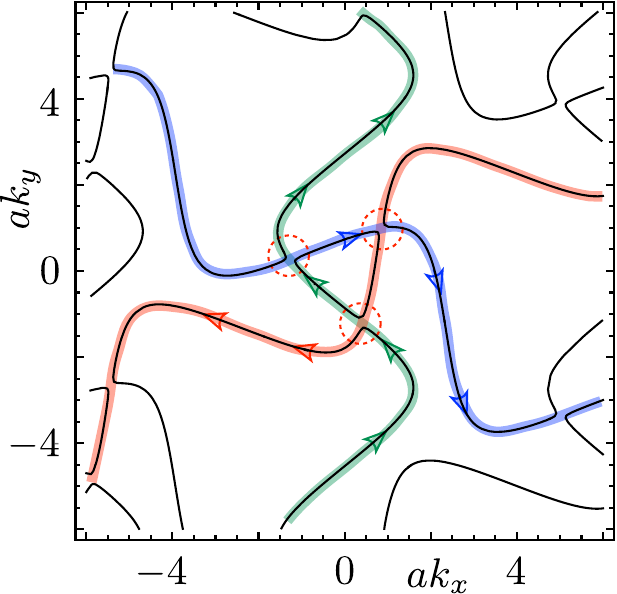}}
\caption{Equi-energy contours $E(k_x,k_y)=0$ at zero magnetic field, computed \cite{note3} for $P_{f_1}=P_{f_2}=1/4$, $P_{d1}=P_{d2}=1/8$, $\alpha=0.1$. A magnetic field drives a wave packet in the direction of the arrows. Points of magnetic breakdown (tunneling between two contours) are encircled. The resulting open orbits are distinguished by different colors.
}
\label{fig_contours}
\end{figure}

We calculate the equi-energy contours \cite{note3} from the scattering matrix \eqref{Sdef}, see Fig.\ \ref{fig_contours}. At $\alpha=0$ three oscillating contours rotated by $120^\circ$ cross near $\bm{k}=0$. A wave packet moves along these open orbits with velocity $dk/dt=v/l_m^2$. A nonzero $\alpha$ opens up a gap $\Delta k\simeq \alpha/a$ at each crossing, thereby allowing the wave packet to be deflected by $\pm 120^\circ$. Magnetic breakdown refers to the tunneling of the wave packet through the gap \cite{Sho84,Ale18}. This happens with the Landau-Zener probability $T=\exp[-c(l_m\Delta k)^2]$, where $c$ is a coefficient of order unity \cite{note4}. We conclude that the breathing mode remains predominantly uncoupled from the side branches provided that $(\alpha l_m/a)^2\ll 1\Rightarrow\alpha^2\ll\Phi/\Phi_0$.

All of this is for the case of equal intra-channel and inter-channel probabilities. We have investigated numerically what happens if we relax this assumption. A difference between $P_{d1}$ and $P_{d2}$ increases the gap, $(a\Delta k)^2\approx\alpha^2+ (P_{d1}-P_{d2})^2$. A difference between $P_{f1}$ and $P_{f2}$ has no effect on the gap, it weakly affects the coefficient $c$.

\textit{Conductance ---} The breathing mode visualized in Fig.\ \ref{fig_breathing} can be detected via the conductance, in the geometry of Fig.\ \ref{fig_bilayer}, with source and drain contacts aligned along a domain wall. We have tested this by computer simulation.

\begin{figure}[tb]
\centerline{\includegraphics[width=0.7\linewidth]{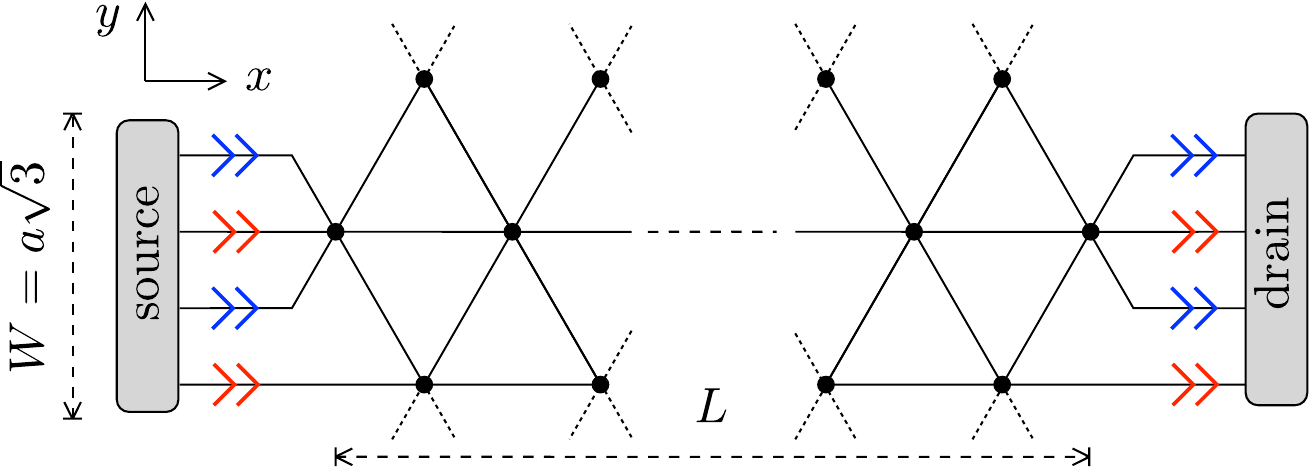}}
\caption{Outgoing modes at the left (source contact) and incoming modes at the right (drain contact). Red and blue arrows distinguish pairs of modes in the two valleys. For contacts of width $W=N\times a\sqrt{3}$ the transmission matrix $t$ from source to drain has dimension $8N\times 8N$. The diagram shows the case $N=1$. The full network in the simulation has length $L$ along the $x$-axis and width much larger than the contact width $W$ along the $y$-axis.
}
\label{fig_leads}
\end{figure}

The transmission matrix $t_{nm}$ from mode $m$ in the source contact to mode $n$ in the drain contact is calculated in the network model \cite{Beu21a}. There are $8N$ outgoing (incoming) modes in the source (drain) contact, distributed over $N=W/a\sqrt 3$ unit cells (see Fig.\ \ref{fig_leads}). Four of the eight modes per unit cell are in one valley and four are in the other valley. 

The two-terminal conductance follows from
\begin{equation}
G=G_0\textstyle{\sum_{n,m=1}^{8N}}|t_{nm}|^2,
\end{equation}
with $G_0=2e^2/h$ the conductance quantum (the factor of two accounts for the spin). In the quasi-1D regime only two of the eight modes per unit cell contribute to the conductance, corresponding to the breathing mode. Note that the current is highly valley polarized: the red modes in Fig.\ \ref{fig_leads} give a negligible contribution to $G$, because they are backscattered into the source at the nodes. A rotation of the contact alignment by $60^\circ$ switches the transmission from one valley to the other.

\begin{figure}[tb]
\centerline{\includegraphics[width=1\linewidth]{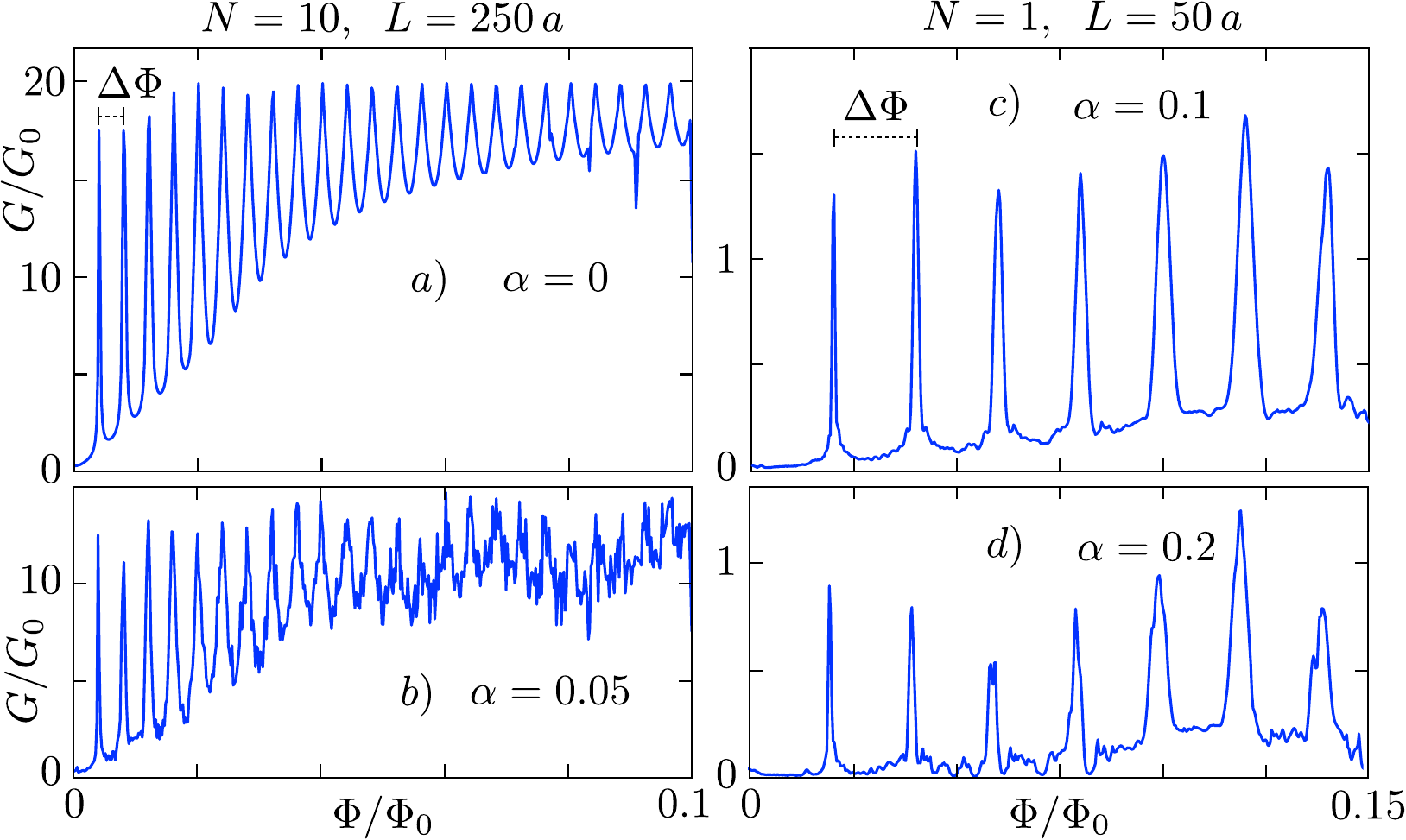}}
\caption{Calculation of the magnetic field dependence of the conductance in the geometry of Fig.\ \ref{fig_bilayer}. Source and drain contacts are separated by $L$ and have a width of $N$ unit cells ($W=N\times a\sqrt{3}$). The parameters of the network model are the same as in Fig.\ \ref{fig_breathing}. Different values of $\alpha$ are compared, for wide contacts (panels a,b) and narrow contacts (panels c,d). The Bloch oscillation period $\Delta\Phi$ from Eq.\ \eqref{deltaphidef} is indicated. Full transmission of the breathing mode corresponds to $G/G_0=2N$. 
}
\label{fig_conductance}
\end{figure}

The conductance is a maximum whenever a node of the breathing mode coincides with the drain contact, so if the separation $L$ of source and drain is an integer multiple of $\pi a/\phi=a\Phi_0/\Phi$. As a function of magnetic field the conductance then oscillates with period \cite{note5}
\begin{equation}
\Delta\Phi=\Phi_0\times a/L\Rightarrow \Delta B=(h/e)(a_yL)^{-1}.\label{deltaphidef}
\end{equation}

This is what we observe in the computer simulation \cite{zenodo}, see Fig.\ \ref{fig_conductance}. To resolve the Bloch oscillations the width $W$ of source and drain contacts should be smaller than the amplitude ${\cal A}\propto 1/B$ of the breathing mode, which explains why the oscillations die out with increasing magnetic field. The oscillations become more robust to nonzero $\alpha$ if both the width and the separation of the contacts are reduced, because then the larger magnetic field scale promotes the magnetic breakdown that enables the breathing mode.

\textit{Discussion ---} In closing, we have shown that the breathing mode that is the hallmark of Bloch oscillations in a periodic potential can be observed in the magnetoconductance of minimally twisted bilayer graphene. The spatial resolution that is needed to resolve the oscillatory electron motion requires narrow source and drain contacts, which is presumably why these low-field oscillations have not yet been observed. Panels c,d in Fig.\ \ref{fig_conductance} correspond to a contact width $W=a\sqrt{3}\approx 0.25\,\mu\text{m}$ at a twist angle $\theta\approx 0.1^\circ$. 

For a contact separation of $L=50\,a\approx 7\,\mu{\rm m}$ the periodicity of the magnetoconductance oscillations is $\Delta B\approx 2.4\,\text{mT}$. This is two orders of magnitude below the fields at which quantum Hall interferometry (Aharonov-Bohm and Shubnikov-De Haas oscillations) becomes operative \cite{Ric18,Xu19,Beu20,Mah21}. There is room to reduce the contact separation, which will help to mitigate disorder effects --- $L$ should be shorter than the mean free path.

The key requirement for the appearance of the breathing mode is the quasi-1D regime, in which open orbits govern the magnetoconductance, enabled by magnetic breakdown. Support for this regime can be found in microscopic calculations of the band structure \cite{Fle20}, that show equi-energy contours qualitatively similar to those in Fig.\ \ref{fig_contours}. The observation of the low-field magnetoconductance oscillations predicted here would then be a striking demonstration of Bloch oscillations in the solid state.

\textit{Acknowledgements ---} We have benefited from discussions with A. R. Akhmerov. This project has received funding from the Netherlands Organization for Scientific Research (NWO/OCW) and from the European Research Council (ERC) under the European Union's Horizon 2020 research and innovation programme.


\begin{thebibliography}{99}
\bibitem{note1} The unitary transformation $S\mapsto USU^\dagger$ that decouples the $6\times 6$ scattering matrix \eqref{Sdef} into $2\times 2$ submatrices is $U=e^{-i(\pi/4)\sigma_y}e^{i(\alpha/2)\sigma_z}$. The Pauli matrices act on the two modes in each domain wall. The decoupling needs $\alpha=0$ and $P_{d1}=P_{d2}$; if $P_{f1}\neq P_{f2}$ the phase $\pi/4$ in Eq.\ \eqref{Rfdef} is replaced by $\arctan\sqrt{P_{f1}/P_{f2}}$.
\bibitem{note2} The constant $y$-component of the vector potential $\bm{A}=(-By,Ba/4,0)$ is chosen such the resulting phase shift does not depend on the parity of the chiral mode. We need this for the mapping to an electric field induced phase shift. All results are obviously gauge independent, but this gauge simplifies the analysis.
\bibitem{note3} The bandstructure $E(\bm{k})$ of the infinite network, as a function of the Bloch momentum $\bm{k}=(k_x,k_y)$, is given by the eigenvalues $e^{iEa/\hbar v}$ of the matrix product $\Omega=S\cdot\text{diag}\,(e^{i\bm{k}\cdot\bm{l}_+},e^{i\bm{k}\cdot\bm{l}_-},e^{i\bm{k}\cdot\bm{l}_0},e^{i\bm{k}\cdot\bm{l}_+},e^{i\bm{k}\cdot\bm{l}_-},e^{i\bm{k}\cdot\bm{l}_0})$, with lattice vectors $\bm{l}_\pm=\tfrac{1}{2}a(1,\pm\sqrt{3})$, $\bm{l}_0=(-a,0)$.
\bibitem{note4} The coefficient $c=\tfrac{1}{4}\pi\tan(\vartheta/2)$ in the magnetic breakdown probability $T=e^{-c(l_m\Delta k)^2}$ is determined by the angle $\vartheta\in(0,\pi)$ at which the equi-energy contours approach the avoided crossing, dependent on the relative direction of motion:
\begin{tikzpicture}[baseline=-0.5ex,scale=.5]
\node (O) at (0,0) {$$};
\node (A) at (2,0.5) {$$};
\node (B) at (2,-0.5) {$$};
\node (C) at (-2,0.5) {$$};
\node (D) at (-2,-0.5) {$$};
\draw [->] (B) -- (O);
\draw [->] (O) -- (C);
\draw [->] (O) -- (A);
\draw [->] (D) -- (O);
\draw[dashed] (A) -- (D);
\draw[dashed] (B) -- (C);
\node (X) at (1.6,0) {$)\vartheta$};
\end{tikzpicture}
or
\begin{tikzpicture}[baseline=-0.5ex,scale=.5]
\node (O) at (0,0) {$$};
\node (A) at (2,0.5) {$$};
\node (B) at (2,-0.5) {$$};
\node (C) at (-2,0.5) {$$};
\node (D) at (-2,-0.5) {$$};
\draw [->] (O) -- (B);
\draw [->] (C) -- (O);
\draw [->] (O) -- (A);
\draw [->] (D) -- (O);
\draw[dashed] (D) -- (A);
\draw[dashed] (C) -- (B);
\node (Y) at (0,0.4) {$\vartheta$};
\draw (1cm,0.4cm) arc (60:120:2cm);
\end{tikzpicture}
For typical parameters we have $\vartheta\approx 60^\circ\Rightarrow c\approx 0.45$.
\bibitem{note5} The periodicity \eqref{deltaphidef} refers to the magnetoconductance oscillations from the breathing mode near $\Phi=0$. Because of gauge invariance, copies exist near higher fields $\Phi_p=\Phi_0/p$ ($p=1,2,\ldots$), with smaller periodicity $\Delta\Phi_p=(\Phi_0/p)\times a/L$.
\end{thebibliography}
\end{document}